\documentclass[manuscript,nonacm]{acmart}

\begin{document}

\title[VoiceGuard]{VoiceGuard: An Audio-Native LLM--VLA Pipeline for Wheelchair-Mounted Assistive Manipulation}

\author{Bang Xiao}
\affiliation{\institution{NYU Shanghai}\city{Shanghai}\country{China}}
\email{bx2181@nyu.edu}
\author{Rodolfo Cossovich}
\affiliation{\institution{Carleton University}\city{Ottawa}\country{Canada}}
\email{rodolfocossovich@cmail.carleton.ca}

\begin{abstract}
Wheelchair-mounted robotic arms can help people with limited upper-limb mobility reach and manipulate everyday objects, but operating them typically requires stepwise joystick control and frequent mode switching. Natural-language interfaces could let users express task-level goals, yet they must handle ambiguous requests, distinguish commands from nearby conversation, and support interruption. We present VoiceGuard, a voice interface that separates spoken intent from robot execution through structured request validation and a local stop mechanism. On-device detectors handle the wake phrase and stop commands, an audio-native large language model (LLM) routes each request into a structured function packet, a rule-based ToolDispatcher accepts, rejects, or clarifies the packet, and a SmolVLA vision-language-action (VLA) policy executes accepted tasks on a low-cost SO-101 arm. Preliminary tests showed that the system correctly handled 55 of 60 voice commands, including ambiguous and non-directed utterances, completed 16 of 20 deliveries, and stopped the arm on all 20 stop commands with a median latency of 210~ms. We demonstrate the feasibility of combining flexible spoken input with constrained task execution and local interruption for wheelchair-mounted assistive manipulation.
\end{abstract}



\begin{teaserfigure}
    \centering
    \includegraphics[width=0.8\textwidth, alt={Pipeline diagram with six components in a row, left to right, grouped as local, cloud, and local. Local Gate (on device): a wake-phrase detector that opens the audio stream and an always-on stop/pause detector, above a photo of a person at the wheelchair setup. Router (cloud): server-side voice activity detection for the start and end of speech, an audio-native LLM, and a function call with a fixed schema. Packet: six fields, namely function, target, destination, module, checks, and transcript. ToolDispatcher (local): packet validation with five ordered checks, leading to four outcomes: Accepted, Clarify, Rejected, and Logged. Accepted packets go to SmolVLA, a VLA policy that takes the instruction and camera images and outputs action chunks, and then to the ArmController, which sends LeRobot move, grasp, release, and home commands to the SO-101 arm and has a stop command. An execution-state arrow returns from the ArmController to the ToolDispatcher. A red arrow, labeled as the stop/pause fast path, runs from the stop/pause detector over the top of the figure directly to the arm's stop, never passing through the cloud or the ToolDispatcher. Below, a Packet example panel shows function deliver, target cup, destination mouth pose, module smolvla cup mouth pose, checks target visible and controller ready, and the transcript "bring the cup up so I can drink". A second panel lists the ToolDispatcher checks in order: (1) schema, function and fields match the schema; (2) arguments, target and destination present and unambiguous; (3) module, a trained module exists and the task is authorized; (4) state, controller ready, target visible, and no interruption; (5) limits on step size, speed, and reach.}]{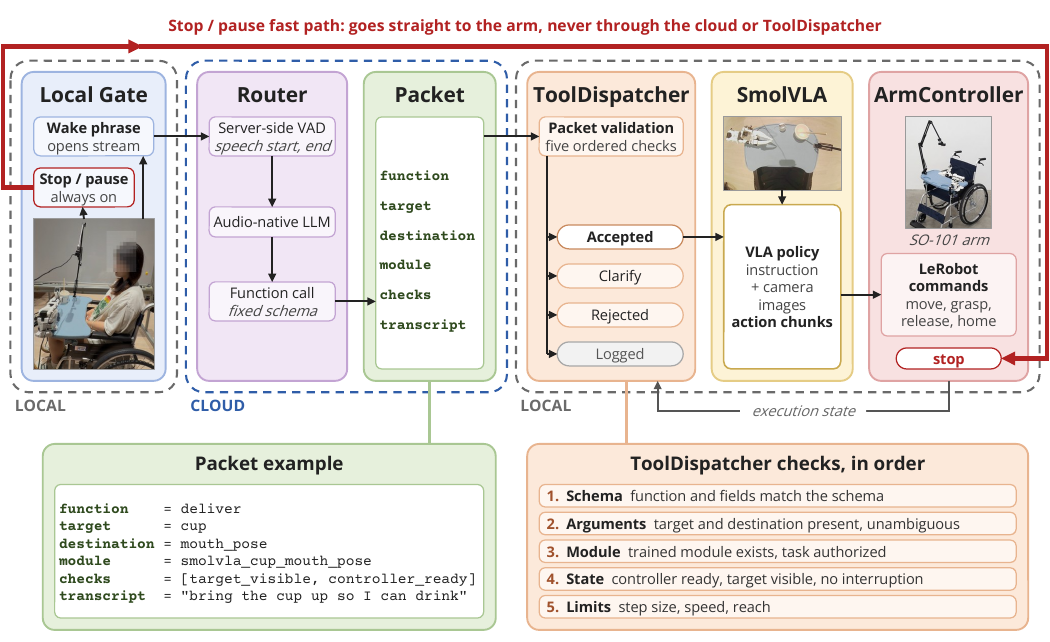}
    \Description{Pipeline diagram with six components in a row, left to right, grouped as local, cloud, and local. Local Gate (on device): a wake-phrase detector that opens the audio stream and an always-on stop/pause detector, above a photo of a person at the wheelchair setup. Router (cloud): server-side voice activity detection for the start and end of speech, an audio-native LLM, and a function call with a fixed schema. Packet: six fields, namely function, target, destination, module, checks, and transcript. ToolDispatcher (local): packet validation with five ordered checks, leading to four outcomes: Accepted, Clarify, Rejected, and Logged. Accepted packets go to SmolVLA, a VLA policy that takes the instruction and camera images and outputs action chunks, and then to the ArmController, which sends LeRobot move, grasp, release, and home commands to the SO-101 arm and has a stop command. An execution-state arrow returns from the ArmController to the ToolDispatcher. A red arrow, labeled as the stop/pause fast path, runs from the stop/pause detector over the top of the figure directly to the arm's stop, never passing through the cloud or the ToolDispatcher. Below, a Packet example panel shows function deliver, target cup, destination mouth pose, module smolvla cup mouth pose, checks target visible and controller ready, and the transcript "bring the cup up so I can drink". A second panel lists the ToolDispatcher checks in order: (1) schema, function and fields match the schema; (2) arguments, target and destination present and unambiguous; (3) module, a trained module exists and the task is authorized; (4) state, controller ready, target visible, and no interruption; (5) limits on step size, speed, and reach.}
    \caption{Pipeline overview. After the wake phrase, the Router turns the request into a Packet, the ToolDispatcher checks it, and SmolVLA and the ArmController carry it out. At any moment, ``stop'' at the Local Gate skips this path and halts the arm.}
    \label{fig:pipeline}
\end{teaserfigure}

\maketitle

%% PDF metadata.
\hypersetup{
  pdftitle={VoiceGuard: An Audio-Native LLM-VLA Pipeline for Wheelchair-Mounted Assistive Manipulation},
  pdfauthor={Bang Xiao, Rodolfo Cossovich},
  pdfdisplaydoctitle=true
}

% =====================================================================
\section{Introduction}
% =====================================================================
    Wheelchair-mounted robotic arms help people with limited upper-limb mobility reach and grasp nearby objects~\cite{rahman2025wheelchair}. While this technology has the potential to reduce the caregiving time needed for daily tasks, their user interfaces are typically driven through a joystick one axis and one mode at a time~\cite{maheu2011evaluation}. Even performing a simple task like picking up a cup requires the users to steer the gripper along each axis, switch modes to rotate the wrist, and switch again to close the gripper.

    With the introduction of Large Language Models (LLMs) and Vision-Language-Action (VLA) Models, task-level control makes robotic arms control more feasible~\cite{liu2026dialogue,shukor2025smolvla}.
    In assistive use, however, the arm moves toward the user's body while the user often cannot physically intervene~\cite{vasic2013safety,adesiji2025safety}, and speech from caregivers reaches the microphone as easily as the user's.
    A voice interface for this setting therefore needs a checkable step between language and motion, and a way to stop the arm at any moment.

    We present VoiceGuard, a voice-controlled system for a wheelchair-mounted robotic arm that validates every request before execution. This paper makes three contributions: (1) an interface architecture design that separates spoken intent from arm motion through an on-device stop path and a validated, logged packet; (2) a working prototype integrating the proposed architecture on a low-cost SO-101 arm mounted on a wheelchair; and (3) preliminary tests of routing, delivery, and stopping performance.

% =====================================================================
\section{Related Work}
% =====================================================================
    Research with people with physical disabilities
    has shown that retrieving everyday objects ranks among their highest priorities in their daily life activities~\cite{anderson2004targeting,choi2009list,hutmacher2025identification}. Commercial
    wheelchair-mounted arms
    address their interface using a joystick~\cite{kinova_jaco,rahman2025wheelchair}.

\textbf{Intelligent Interfaces for assistive arms: }
    Existing interfaces reduce the burden of joystick control  through shared autonomy, target selection, and speech. Shared-autonomy systems automate mode switching~\cite{herlant2016assistive} or infer the user's goal~\cite{jain2019probabilistic}, but keep the user on the joystick. Selection-based systems let users pick a target by eye gaze or pointing~\cite{abayasiri2024intelligent,choi2009handitover,jain2010ele,muraleedharan2024eyegaze}. LLM-based speech interfaces for feeding~\cite{yuan2024towards} and joint wheelchair and arm control~\cite{liu2026dialogue} accept free-form requests. Our system differs by placing validation and stopping outside the model.

\textbf{Language and VLA models for robot action: }
    Technically, our work builds on LLM task planning and VLA policies. SayCan, Code as Policies, and ProgPrompt use LLMs to select or program robot tasks~\cite{ahn2023saycan,liang2023code,singh2023progprompt}, and VLA models such as RT-2, OpenVLA, and SmolVLA map an instruction and camera images to actions~\cite{brohan2023rt2,kim2024openvla,shukor2025smolvla}. These approaches were mostly evaluated in tabletop settings, where model output is executed directly. We adapt them for assistive use by restricting the LLM to a fixed function schema and giving each VLA policy only the task instruction it was trained on.

\textbf{Safe motion near the user: }
    Robots that work next to people rely on low-level protections such as collision detection from joint torque or motor current and limits on speed, force, and power~\cite{haddadin2017robot,villani2018survey}. These protections respond to contact or to exceeded limits, and do not capture what the user wants the arm to do. Our work adds a software layer above them: the ToolDispatcher bounds each task before the arm moves, and the local stop lets the user halt motion by voice at any point.

% =====================================================================
\section{System Design}
% =====================================================================
    We designed the system around four goals: to let users phrase task-level requests in their own words (DG1); to check, reject, or clarify every request before the arm moves, and log it (DG2); to keep stopping independent of network or model failures (DG3); and to send audio off the device only after the wake phrase, so that nearby conversations never reach the cloud model (DG4). We realize these goals with six components (Fig.~\ref{fig:pipeline}).
    The local gate listens on the device for the wake phrase and for stop commands. After the wake phrase, the audio-native router turns the spoken request into a structured function packet. The ToolDispatcher checks the packet and decides whether it may run. If it passes, the SmolVLA policy generates arm actions, and the ArmController sends them to the SO-101 arm.

\subsection{Audio-Native Routing}
    The local gate runs two detectors on the device: a wake-phrase detector, which we set to ``Hey Rudi,'' that opens a streaming session to an audio-native LLM (DG4), and a stop and pause detector that stays always on. The LLM, Gemini 3.5 Flash-Lite in our prototype, returns a function call over a fixed schema of supported functions (DG1); any audio-native model with function calling can replace it. We route raw audio instead of a transcript, which removes a separate speech-to-text stage and its errors on short commands.

    The function call becomes a packet with six fields (Fig.~\ref{fig:pipeline}, top right). For ``bring the cup up so I can drink,'' the router fills \texttt{target=cup} and \texttt{destination=mouth\_pose}, names the action module, lists the checks the packet must pass, and records a \texttt{transcript} so that each decision can be audited (DG2). ``Bring that'' yields a clarification request, and a remark addressed to a caregiver yields no packet.

\subsection{Validation}
    The ToolDispatcher applies five checks in order: (1) the function and fields match the schema; (2) the target and destination are present and unambiguous; (3) a trained module exists and the task is authorized; (4) the controller is ready, the target is visible, and no interruption is active; and (5) step size, speed, and reach stay within configured limits. A packet that passes all checks is accepted; a missing or ambiguous argument returns a clarification request, and any other failure is rejected without motion. Every packet and outcome is logged.

\subsection{Execution and Stop}
    An accepted packet is converted into the fixed task instruction its module was trained on, so the policy never receives open-ended speech. Each module is a SmolVLA policy~\cite{shukor2025smolvla} fine-tuned for one target and destination pair. The policy takes the instruction together with wrist and agent camera images and outputs action chunks, which the ArmController converts into LeRobot commands (move, grasp, release, home) for the SO-101 arm. Stop and pause go from the local gate straight to the ArmController, so the arm halts without a network round trip or a model call (DG3).

% =====================================================================
\section{Prototype and Interaction}
\begin{figure}[tb]
    \centering
    \includegraphics[width=1\textwidth, alt={Five photographs in a row, labeled (a) to (e), showing one delivery at a wheelchair tray. A seated person, face blurred, sits beside an SO-101 robotic arm mounted next to a blue tray. (a) Standby pose, awaiting task input: labels mark the wrist camera, the agent camera, and the arm's safe ready pose. (b) Voice input, command received: a speech bubble shows the request "bring me the water". (c) VLA perception, scene grounding: the policy grounds the task in the current camera view. (d) VLA planning, action generation: the policy generates an executable motion plan and the arm moves toward the cup on the tray. (e) Target reached, task completed: the arm brings the cup to the requested position near the person.}]{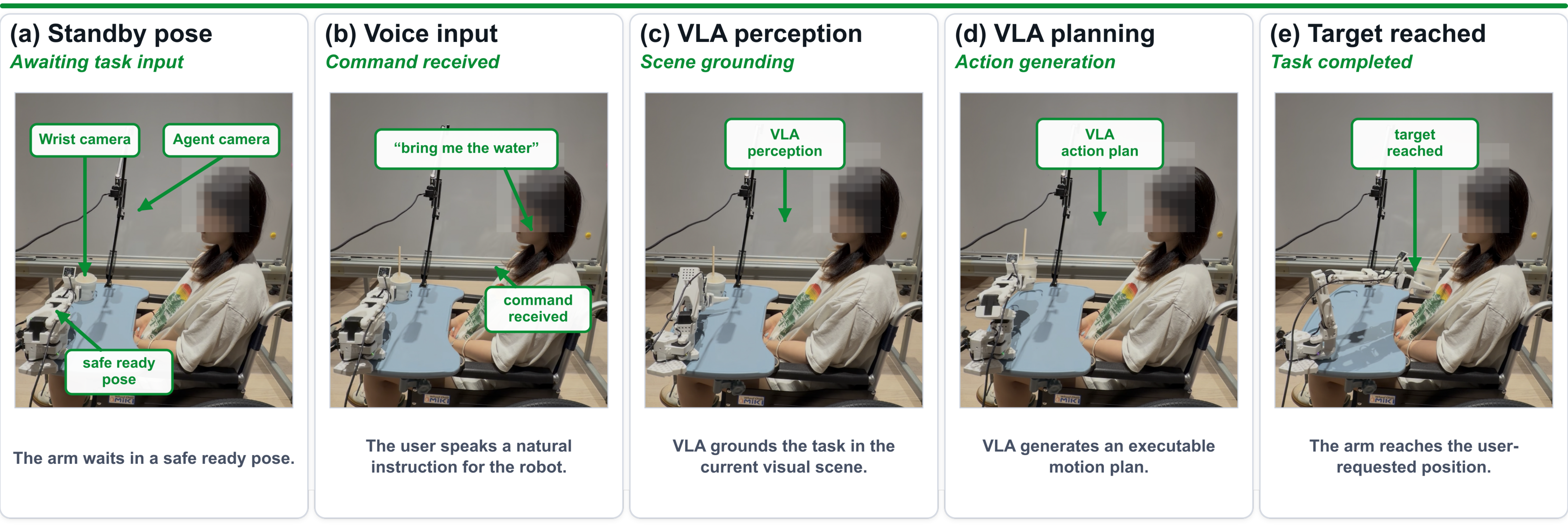}
    \Description{Five photographs in a row, labeled (a) to (e), showing one delivery at a wheelchair tray. A seated person, face blurred, sits beside an SO-101 robotic arm mounted next to a blue tray. (a) Standby pose, awaiting task input: labels mark the wrist camera, the agent camera, and the arm's safe ready pose. (b) Voice input, command received: a speech bubble shows the request "bring me the water". (c) VLA perception, scene grounding: the policy grounds the task in the current camera view. (d) VLA planning, action generation: the policy generates an executable motion plan and the arm moves toward the cup on the tray. (e) Target reached, task completed: the arm brings the cup to the requested position near the person.}
    \caption{A delivery on the physical prototype, from the ready pose (a) to the requested pose (e).}
    \label{fig:workflow}
\end{figure}

% =====================================================================
    The prototype mounts an SO-101 arm beside a wheelchair tray, with a wrist camera and a fixed agent camera (Fig.~\ref{fig:workflow}). The local gate runs openWakeWord on a laptop, and the router uses the Gemini Live API with Gemini 3.5 Flash-Lite. Each SmolVLA module was fine-tuned on 50 teleoperated demonstrations and runs on an RTX 4070 GPU through LeRobot. A delivery takes one utterance (Fig.~\ref{fig:workflow}). ``Stop'' or ``pause'' halts the arm at any point, and ``bring that'' leads to a spoken prompt to name the object.

% =====================================================================
\section{Preliminary Evaluation}
% =====================================================================
    We tested the prototype's three core functions (Table~\ref{tab:prelim}). For command routing, the first author said 10 phrasings for each of four supported requests plus 20 ambiguous or non-directed utterances; a command counted as correct when it produced the expected packet, clarification request, or no packet. Deliveries used the setup in Fig.~\ref{fig:workflow}. For stopping, we used the word ``stop'' while the arm approached the target, and we measured latency from the end of the utterance to the arm's last joint motion. Most failed deliveries were missed grasps, and delivery success varied with lighting and background, as reported for other VLA policies~\cite{fei2025liberoplus,pumacay2024colosseum}.

\begin{table}[tb]
    \caption{Preliminary tests on the physical prototype.}
    \label{tab:prelim}
    \small
    \begin{tabular}{@{}lll@{}}
        \toprule
        Function & Trials & Result \\
        \midrule
        Command routing    & 60 & 55/60 correct packet, clarification, or no packet \\
        Physical execution & 20 & 16/20 deliveries completed \\
        Stop mechanism     & 20 & 20/20 halted; median latency 210~ms \\
        \bottomrule
    \end{tabular}
\end{table}

\section{Conclusion and Future Work}
    We presented VoiceGuard, a voice interface for a wheelchair-mounted robotic arm that turns spoken requests into validated packets and keeps stop on the device; in our prototype, it halted the arm on every stop command. More broadly, by treating language-model output as a proposal to check before execution, such systems can help assistive robotics move from step-by-step teleoperation toward task-level requests the user still controls.

    Our tests involved lab members running trained routines on a fixed tray setup. A study with wheelchair users is now under ethics review and will examine how they phrase requests and respond to clarification prompts. We also plan to run routing on the device, removing the need for a network connection, and to improve the robustness of the VLA modules to the lighting and background changes that lowered delivery success in our tests.

\bibliographystyle{ACM-Reference-Format}
\bibliography{refs}

\end{document}